\documentclass[aps,apl,twocolumn]{revtex4-1}
\usepackage{graphicx}
\usepackage{dcolumn}
\usepackage{bm}
\newcommand{\mb}{\boldsymbol}
\begin{document}
\preprint{}

\title{Long-lived nanosecond spin relaxation and spin coherence of electrons in monolayer MoS$_2$ and WS$_2$}

\author{Luyi Yang$^1$, Nikolai A. Sinitsyn$^2$, Weibing Chen$^3$, Jiangtan Yuan$^3$, Jing Zhang$^3$, Jun Lou$^3$, Scott A. Crooker$^1$,}

\affiliation{$^1$National High Magnetic Field Laboratory, Los Alamos, NM 87545}
\affiliation{$^2$Theoretical Division, Los Alamos National Laboratory, Los Alamos, NM 87545}
\affiliation{$^3$Department of Materials Science and NanoEngineering, Rice University, Houston, TX 77005}
 \maketitle
\textbf{The recently-discovered monolayer transition metal dichalcogenides (TMDCs) provide a fertile playground to explore new coupled spin-valley physics \cite{Xiao, XuReview, ValleyHall}. Although robust spin and valley degrees of freedom are inferred from polarized photoluminescence (PL) experiments \cite{Splendiani, Mak2010, Zeng, Sallen, Jones2013}, PL timescales are necessarily constrained by short-lived (3-100ps) electron-hole recombination \cite{Lagarde, WangG}. Direct probes of spin/valley polarization dynamics of \emph{resident} carriers in electron (or hole) doped TMDCs, which may persist long after recombination ceases, are at an early stage \cite{Zhu, Plechinger, DalConte}. Here we directly measure the coupled spin-valley dynamics in electron-doped MoS$_2$ and WS$_2$ monolayers using optical Kerr spectroscopy, and unambiguously reveal very long electron spin lifetimes exceeding 3ns at 5K (2-3 orders of magnitude longer than typical exciton recombination times). In contrast with conventional III-V or II-VI semiconductors, spin relaxation accelerates rapidly in small transverse magnetic fields. Supported by a model of coupled spin-valley dynamics, these results indicate a novel mechanism of itinerant electron spin dephasing in the rapidly-fluctuating internal spin-orbit field in TMDCs, driven by fast intervalley scattering. Additionally, a long-lived spin \emph{coherence} is observed at lower energies, commensurate with localized states. These studies provide crucial insight into the physics underpinning spin and valley dynamics of resident electrons in atomically-thin TMDCs.}

Studies of optical spin orientation and spin relaxation using polarized light have a long and exciting history in conventional III-V and II-VI semiconductors \cite{DP, AwschalomBook}.  Early seminal works focused on magneto-optical studies of polarized PL from recombining excitons \cite{DP}, from which spin lifetimes could be indirectly inferred. However, it was the direct observation of very long-lived spin coherence of resident electrons in materials like GaAs and ZnSe \cite{KikkawaPRL, KikkawaScience} -- revealed unambiguously by time-resolved Faraday and Kerr rotation studies -- that captured widespread interest and helped to launch the burgeoning field of ``semiconductor spintronics" in the late 1990s \cite{AwschalomBook}. With a view towards exploring coupled spin/valley physics of resident electrons in the new atomically-thin and direct-bandgap TMDC semiconductors, here we apply related experimental methods and directly reveal surprisingly long-lived and coherent spin dynamics in monolayer MoS$_2$ and WS$_2$.

Figure 1a depicts the experimental setup. High-quality monolayer crystals of \emph{n}-type MoS$_2$ and WS$_2$, grown by chemical vapor deposition on SiO$_2$/Si substrates \cite{JLou}, were selected based on low-temperature reflectance and PL studies (see Methods). Transverse magnetic fields ($B_y$) were applied using external coils. A weak pump laser illuminates individual crystals with right- or left-circularly polarized light (RCP or LCP) using wavelengths near the lowest-energy ``A" exciton transition, which primarily photoexcites spin-polarized electrons and holes into the $K$ or $K'$ valley, respectively \cite{Xiao, XuReview, ValleyHall, Splendiani, Mak2010, Zeng, Sallen, Jones2013, Lagarde, WangG}. Any induced spin and valley polarization is then detected via the optical Kerr rotation (KR) or Kerr ellipticity (KE) that is imparted to a linearly-polarized and wavelength-tunable probe laser that is normally incident and focused on to the crystal. Either continuous-wave (cw) or pulsed pump/probe lasers can be used; both types of experiments will be discussed.

\begin{figure*}[tbp]
\includegraphics[width=.82\textwidth]{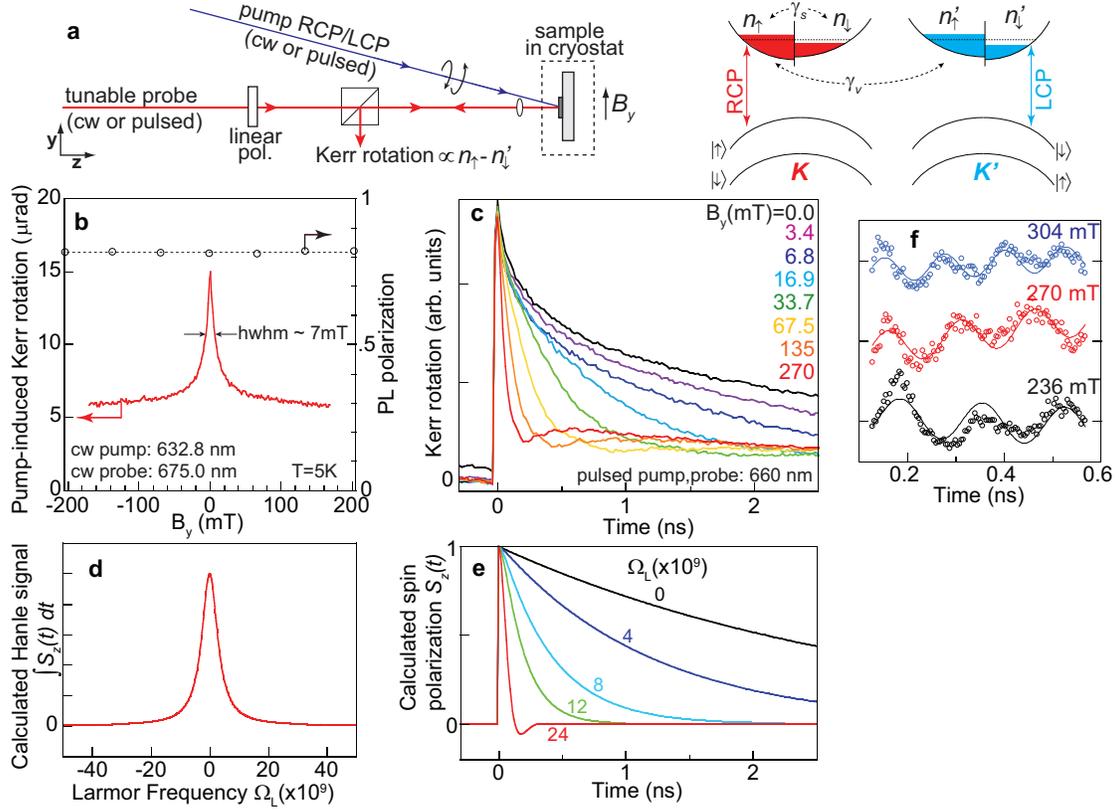}
\caption{\textbf{Long-lived electron spin dynamics in n-type MoS$_2$ at 5K.} \textbf{a}, Experimental schematic. A weak pump laser illuminates the TMDC crystals with right- or left-circularly polarized (RCP or LCP) light. The induced spin/valley polarization is detected via the optical Kerr rotation (KR) or Kerr ellipticity (KE) that is imparted on a linearly-polarized, wavelength-tunable probe laser.  Pump and probe lasers can be either continuous-wave (cw) or pulsed.  The diagrams depict a simple single-electron picture of the conduction and valence bands at the $K$ and $K'$ valleys of electron-doped monolayer MoS$_2$, along with the relevant optical selection rules and scattering processes. In each valley, the conduction bands are separately drawn as spin-up (left) and spin-down (right) components (with slightly different curvature and with small splitting $\Delta_c$; these effects generate the effective spin-orbit field $\pm \hat z B_{\textrm{so}}$). The densities of the resident electrons are $n_\uparrow$, $n_\downarrow$, $n'_\uparrow$, and $n'_\downarrow$. Near the lowest-energy ``A" exciton, RCP light couples only to spin-up electron states in the $K$ valley ($n_\uparrow$), while LCP light couples only to spin-down electron states in the $K'$ valley ($n'_\downarrow$). Densities $n_\downarrow$ and $n'_\uparrow$ do not couple directly to light at the A exciton energy. Following recombination with photogenerated holes, KR and KE therefore depend on the difference $n_\uparrow - n'_\downarrow$. Electron spin relaxation  within a given valley ($\gamma_s$) couples $n_\uparrow$ with $n_\downarrow$ (and $n'_\uparrow$ with $n'_\downarrow$), while spin-conserving intervalley scattering $\gamma_v$ couples $n_\uparrow$ with $n'_\uparrow$ (and $n_\downarrow$ with $n'_\downarrow$). \textbf{b}, Using weak cw pump and probe lasers, the red curve shows the induced KR versus applied magnetic field $B_y$ from monolayer MoS$_2$ at 5~K. This ``Hanle-Kerr" experiment reveals rapid reduction of optically-induced spin polarization by very small $B_y$, suggesting long spin relaxation times of resident electrons. The PL polarization shows no change over this field range. \textbf{c}, Time-resolved KR data (using pulsed pump and probe lasers) directly reveals very long electron relaxation dynamics in MoS$_2$, at different $B_y$. \textbf{d,e}, Calculated Hanle-Kerr and time-resolved dynamics of itinerant resident electrons, based on the model developed in the main text. Fast intervalley scattering $\gamma_v$ and associated fluctuating $\pm \hat z B_{\textrm{so}}$ drives the rapid dephasing of electron spin polarization in small $B_y$. \textbf{f}, Expanded view of residual long-lived electron spin coherence, likely due to additional contributions from localized states (curves offset for clarity).} \label{fig1}
\end{figure*}

The diagrams in Fig.~1a depict the conduction and valence bands in the $K$ and $K'$ valleys in a typical monolayer TMDC, along with the relevant spin/valley optical selection rules and scattering processes. In \emph{n}-type material, the resident spin-up and spin-down electrons in the conduction band of the $K$ ($K'$) valley have densities $n_\uparrow$ and $n_\downarrow$ ($n'_\uparrow$ and $n'_\downarrow$), which all share a common chemical potential in thermal equilibrium. However, following pulsed photoexcitation and fast ($\sim$10~ps) recombination with photogenerated holes \cite{Lagarde, WangG}, these resident electron densities may be \emph{unequal} and \emph{out of equilibrium}, as depicted. This can arise, \emph{e.g.}, from many-body correlations while holes are present \cite{Mai, ElaineLi}, or preferential non-radiative recombination of holes with particular electron states (nonradiative processes account for the majority of recombination in TMDCs \cite{Mak2010}). In this way, photoexcitation can impart a net spin polarization ($S_z=n_\uparrow - n_\downarrow + n'_\uparrow - n'_\downarrow$) and/or valley polarization ($N_v=n_\uparrow + n_\downarrow - n'_\uparrow - n'_\downarrow$) onto the resident electrons, that may remain even after all holes have recombined. (Analogously, a weak steady-state photoexcitation can establish a nonequilibrium steady-state polarization of resident electrons). The intrinsic relaxation dynamics of this polarization, which proceeds without the perturbing influence of the holes, is the principal focus of these studies.

Crucially, any long-lived polarization of the resident electrons can be directly monitored using Kerr spectroscopy.  This stands in marked contrast with polarized PL studies, which explicitly require the participation of (and recombination with) a photo-excited hole. Kerr effects depend only on the difference between a material's RCP and LCP absorption and indices of refraction. Per the selection rules in monolayer MoS$_2$ and related TMDCs \cite{Xiao, XuReview}, RCP light near the lowest-energy ``A" exciton in TMDCs couples to the resident electron density $n_\uparrow$ in the $K$ valley. Similarly, LCP light couples to $n'_\downarrow$ in the $K'$ valley. Nonequilibrium perturbations to the densities $n_\uparrow$ and $n'_\downarrow$ shift their chemical potentials, which change the absorption and refraction of RCP and LCP probe light, particularly at wavelengths near optical transitions. Thus, to leading order, Kerr signals using light near the A exciton are proportional to $n_\uparrow - n'_\downarrow$, to which both spin and valley polarization can contribute, \emph{viz.} $(S_z + N_v)/2$.

Figure 1b demonstrates optically-induced polarization in monolayer MoS$_2$ using cw pump and probe lasers.  The wavelengths $\lambda_{pump}$ (632.8~nm) and $\lambda_{probe}$ (675~nm) were chosen to address only the ``A" exciton. At zero applied field, a steady-state KR of $\sim$15 $\mu$rad is induced by the pump onto the probe laser. Surprisingly, however, this polarization signal is sharply reduced by very small transverse fields $B_y$. The very narrow and nearly Lorentzian-shaped dependence of the measured KR on $B_y$ (the half-width is only $\sim$7~mT) strongly suggests a long-lived spin polarization of resident electrons. (In contrast, the PL polarization, which probes primarily short-lived \emph{excitons}, is unchanged over this field range, in agreement with earlier studies \cite{Sallen}). These Kerr data are very reminiscent of traditional ``Hanle-effect" studies in conventional semiconductors like \emph{n}-type GaAs \cite{DP, Dzhioev, Furis}, wherein $B_y$ dephases an optically injected steady-state electron spin polarization ($S_z$) due to spin precession about $B_y$. This leads to a Lorentzian dependence of $S_z$ on $B_y$, from which the spin lifetime $\tau_s$ can be directly inferred via the half-width $B_{y0}$ of the Hanle peak if the g-factor $g_e$ is known; namely, $\tau_s^{-1} = g_e \mu_B B_{y0} /\hbar$.

It is therefore tempting to associate the data in Fig. 1b with spin dephasing in MoS$_2$ due solely to precession of resident electrons about $B_y$, and to infer a spin lifetime.  However, in contrast to conventional III-V or II-VI semiconductors, electrons in the high-momentum $K$ and $K'$ valleys of monolayer MoS$_2$ are predicted to experience strong spin-orbit coupling, due to the different curvature of the spin-up and spin-down conduction bands and to any intrinsic splitting $\Delta_c$ of these bands \cite{Xiao, XuReview, Kormanyos, Tse}.  This coupling can be viewed as a large (of order 10~T) out-of-plane effective magnetic field $\textbf{B}_{\textrm{so}}$.  Importantly, $\textbf{B}_{\textrm{so}}$ is oriented parallel or antiparallel to $\hat z$ depending on whether the electron resides in the $K$ or $K'$ valley. Spin-conserving intervalley electron scattering, which is not forbidden in the conduction band and is expected to be fast ($\gamma_v^{-1} \sim$0.1-1ps), therefore leads to a rapidly fluctuating effective magnetic field `seen' by electrons.

This fluctuating field alone will not affect (dephase) electron spins that are also oriented along $\pm \hat z$.  However, in the additional presence of $B_y$, electron spins \emph{will} precess about the total fluctuating field $\hat y B_y  \pm \hat z B_{\textrm{so}}$, which is no longer oriented along $\hat z$.  This leads to a valley-dependent spin precession and associated spin dephasing that is analogous to momentum-dependent spin precession and dephasing common in conventional semiconductors (\emph{e.g.}, the Dyakonov-Perel mechanism \cite{DP}) or to the electron spin depolarization in germanium that is also driven by intervalley scattering \cite{DeryGe}.  A \emph{direct and testable consequence} is that electron spin relaxation in MoS$_2$ is expected to depend strongly on $B_y$ -- even for small values of $B_y$ -- in marked contrast to ordinary III-V and II-VI bulk semiconductors.

\begin{figure}[tbp]
\includegraphics[width=.45\textwidth]{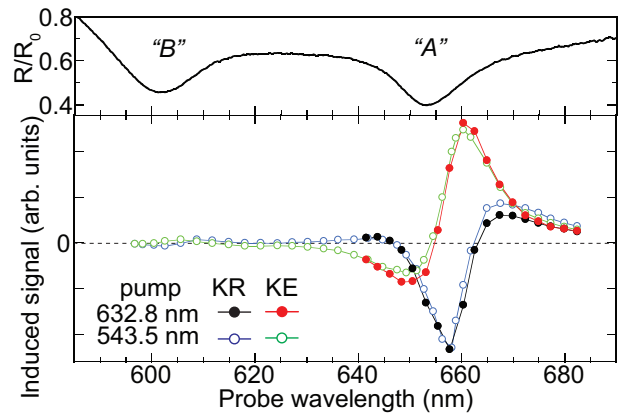}
\caption{{\bf Spectral dependence of the optically-induced Kerr rotation/ellipticity signals in monolayer MoS$_2$}.  The upper (black) trace shows the normalized reflectance spectrum $R/R_0$ from a MoS$_2$ crystal at 5~K.  ``A" and ``B" exciton features are clear. The lower traces show the optically-induced KR and KE signals as a function of the probe laser wavelength at $B_y$=0. A strong resonance at the ``A" exciton is observed.  Results using two different pump laser wavelengths are shown (632.8 nm and 543.5 nm)} \label{fig2}
\end{figure}

The narrow Hanle-Kerr data in Fig. 1b are consistent with this scenario.  However, to \emph{directly} measure electron spin relaxation and test this hypothesis, we turn to time-resolved KR studies using ultrafast pump and probe lasers. Figure 1c shows the measured decay of the optically-induced electron polarization in the same MoS$_2$ crystal, using photons with energy just below the ``A" exciton resonance. A key finding of this work is that the polarization decay time is extremely long ($\sim$3~ns) at zero field. This timescale exceeds typical PL recombination times by two to three orders of magnitude \cite{Lagarde, WangG}, further implicating resident electrons as the source of long-lived polarization. Very recent studies of monolayer MoS$_2$ and WSe$_2$ using related optical techniques did not reveal any longlived polarization imparted to resident electrons, perhaps due to elevated temperatures \cite{DalConte} or to the use of probe photons with higher energy \cite{Zhu, Plechinger}. The nanosecond electron spin relaxation times observed here at zero field may ultimately be limited by intra-valley Dyakonov-Perel or Elliot-Yafet processes \cite{Ochoa, WangL}, or spin-flip scattering with magnetic impurities or long-wavelength flexural phonons \cite{Dery}.

Crucially, the polarization decay time rapidly decreases with small increasing $B_y$, and no prominent spin precession is observed. These two observations directly support the scenario described above, of depolarization due to a rapidly-fluctuating $\textbf{B}_{\textrm{so}}$ driven by fast inter-valley electron scattering, for the following reasons: i) If $\textbf{B}_{\textrm{so}}$ were small or did not exist, then pronounced spin precession about $B_y$ would be observed (as is the case for resident electrons in GaAs, ZnSe, GaN, or CdTe \cite{AwschalomBook, KikkawaScience, KikkawaPRL, Zhukov, Cundiff}), and ii) if $\textbf{B}_{\textrm{so}}$ were static and not fluctuating, then spins would be effectively pinned along $\textbf{B}_{\textrm{so}}$ and the small $B_y$ ($\ll |\textbf{B}_{\textrm{so}}|$) would have little influence on the long spin decay. Neither of these phenomena were observed. Rather, these data are consistent with a novel spin relaxation mechanism in monolayer TMDCs that is driven by fast inter-valley scattering (rapidly fluctuating $\textbf{B}_{\textrm{so}}$), and activated by small $B_y$.

A model of coupled spin-valley dynamics captures the underlying physics and reproduces essential features of the data. The phenomenological equation of motion for the electron spin polarizations $\mb{S}^K$ and $\mb{S}^{K'}$ in the two valleys reads:
\begin{equation}
\label{eq:EOM}
\frac{d\mb{S}^\tau}{dt}=\mb{\Omega_L} \times \mb{S}^\tau + \tau \Omega_{\textrm{so}}\hat{z} \times \mb{S}^\tau
-\gamma_s \mb{S}^\tau - \gamma_v(\mb{S}^\tau - \mb{S}^{-\tau}),
\end{equation}
where $\tau= \tau(t)=\pm 1$ is the index for $K$ and $K'$ valleys respectively. The first two terms describe spin precession about an applied field $\mb{\Omega}_L = g\mu_B\mb{B}/\hbar$ and the internal spin-orbit field $\mb{\Omega}_{\textrm{so}} =\pm \hat z g \mu_B B_{\textrm{so}}/\hbar$. The 3rd term describes intrinsic spin relaxation within a given valley with rate $\gamma_s$, and the 4th term is spin-conserving inter-valley scattering with rate $\gamma_v$ that is expected to be fast and which leads to fast relaxation of any valley polarization.
We compute the total spin polarization $ \mb{S}= \mb{S}^{K} +\mb{S}^{K'} $.
For large $\textbf{B}_{\textrm{so}}$ ($\Omega_{\textrm{so}}\gg\Omega_L$) and fast intervalley scattering ($\gamma_v>\Omega_{\textrm{so}}$), and assuming initial polarization $S_0\hat{z}$, the solution is $S_z(t)=S_0\sum_{j=1,2}A_je^{i\omega_jt}$.  Here, $A_{1,2}=(1\pm\Gamma_v/\sqrt{\Gamma_v^2-\Omega_L^2})/2$ and the two eigenmodes $i\omega_{1,2}=-\gamma_s-\Gamma_v\pm\sqrt{\Gamma_v^2-\Omega_L^2}$, where $\Gamma_v\equiv\Omega_{\textrm{ so}}^2/4\gamma_v$. There is a critical applied field $\Omega_L^{c}=\Gamma_v$, below which the modes are purely decaying and above which the modes oscillate. The Hanle curve is calculated by integrating $S_z(t)$ for each $\Omega_L$; namely $\int_0^{\infty} S_z(t) dt = S_0 (\gamma_s+2\Gamma_v)/[\Omega_L^2+(\gamma_s^2+2\gamma_s\Gamma_v)]$.  Thus, Hanle-Kerr data are predicted to be Lorentzian (as observed) with half-width $\sqrt{\gamma_s^2+2\gamma_s\Gamma_v}$.

Figures 1d and 1e show the calculated Hanle curves and electron spin dynamics. The model reproduces the rapid drop in electron spin polarization with increasing (small) $B_y$, and captures the shallow dip and subsequent recovery of the electron spin polarization at short timescales. The model does not, of course, capture the offset of the measured Hanle-Kerr data (see Fig. 1b), the origin of which suggests an additional long-lived and field-independent polarization, perhaps from localized states in MoS$_2$, which have been studied recently. This offset is also manifested in the time-resolved data as the very slowly decaying and largely field-independent signal that persists at long delays.

\begin{figure}[tbp]
\includegraphics[width=.48\textwidth]{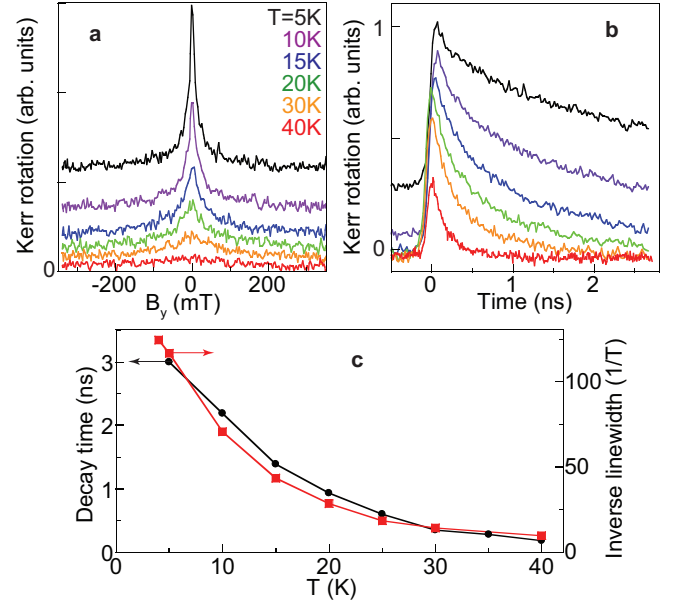}
\caption{{\bf Temperature dependence of electron spin relaxation in monolayer MoS$_2$ at zero magnetic field.}  \textbf{a}, Hanle-Kerr data (using cw lasers; $\lambda_{pump}$=632.8 nm, $\lambda_{probe}$=675~nm). With increasing temperature, the curve width increases while the amplitude drops, indicating faster spin relaxation of resident electrons.  \textbf{b}, Corresponding time-resolved KR directly reveals faster spin relaxation with increasing temperature. Here, $\lambda_{pump}$=635~nm with $\sim$100~ps pulse duration; $\lambda_{probe}$=672~nm with $\sim$200~fs pulse duration. \textbf{c}, The measured relaxation time, and inverse Hanle-Kerr linewidth, versus temperature.} \label{fig4}
\end{figure}

A surprising additional observation is the appearance, at larger magnetic fields, of a small but long-lived oscillatory signal visible between $\sim$200-800 ps (red trace in Fig. 1c).  Figure 1f shows an expanded view of this signal at $B_y$=236, 270, and 304 mT. The oscillation frequency scales linearly with $B_y$, indicating that this signal arises from coherently-precessing electrons with g-factor $|g_e| \simeq 1.8$. Such long-lived coherence signals are not expected from \emph{itinerant} resident electrons for reasons described above. However, they may arise from contributions from an additional population of localized states that do not undergo rapid intervalley scattering and which precess about the bare applied field $B_y$. Time resolved measurements at lower photon energies below the A exciton are consistent with this scenario (Fig. S1).

We confirm that these Kerr signals originate from MoS$_2$ by showing their spectral dependence.  Figure 2 shows the reflectivity of the MoS$_2$ crystal (top black trace), below which are the peak Hanle-Kerr signals at $B_y$=0.  Both KR and KE show a strong resonance at the ``A" exciton. Data using 632.8~nm and 543.5~nm cw pump lasers are shown; the latter allows to track the induced Kerr signals out to shorter probe wavelengths that overlap with the higher-energy ``B" exciton which elicits a smaller response. This behavior was confirmed on many different MoS$_2$ crystals. We note that the resonances are redshifted $\sim$25 meV from the exciton peak, suggesting that the resident electrons' polarization is revealed preferentially through trion-related (rather than neutral exciton-related) optical transitions \cite{XuReview}, consistent with analogous studies in III-V and II-VI semiconductor quantum wells \cite{Zhukov, Cundiff}.  The dependence of Hanle-Kerr data on probe wavelength is shown in Fig. S2.

\begin{figure}[tbp]
\includegraphics[width=.48\textwidth]{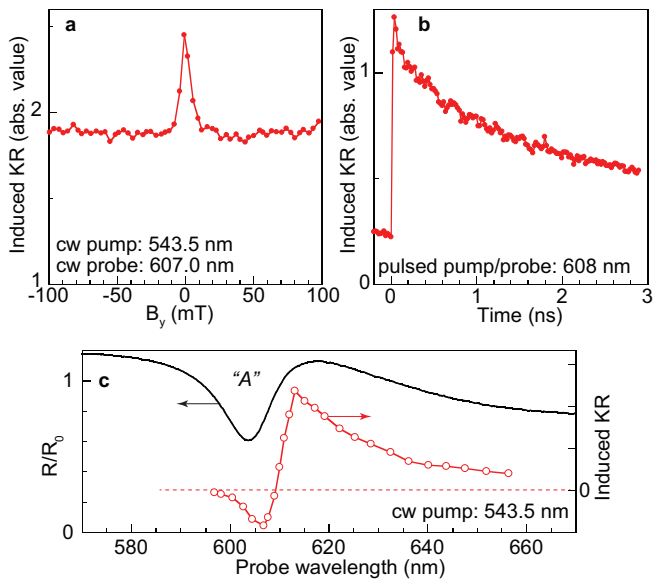}
\caption{{\bf Long-lived spin polarization dynamics in monolayer WS$_2$ at 5~K}. \textbf{a}, Hanle-Kerr measurement of optically-induced spin polarization at the ``A" exciton, using cw lasers. (At this probe wavelength the induced KR has negative sign, thus the absolute value is shown for clarity). \textbf{b}, Time-resolved KR at zero field (using ultrafast pump and probe pulses). \textbf{c}, The normalized reflectivity of monolayer WS$_2$ (left axis), showing the ``A" exciton feature.  Right axis: wavelength dependence of the optically-induced KR signal at zero applied field (using cw lasers).} \label{fig5}
\end{figure}

An essential consideration for spin-coherent phenomena in semiconductors is its dependence on temperature. Systematic studies on MoS$_2$ (see Fig. 3) reveal very long spin lifetimes at 5K ($\sim$3~ns), which decrease rapidly to $<$200~ps by 40~K. This behavior is in qualitative agreement with recent theoretical predictions \cite{Dery} that account for Elliot-Yafet spin-flip processes due to electron-phonon scattering with long-wavelength flexural phonons in TMDCs. A comparison of supported versus unsupported monolayer crystals could confirm this mechanism.

Finally, we show that these long-lived electron spin polarizations are not unique to MoS$_2$, but also appear in other TMDCs. Figure 4 shows Hanle-Kerr and time-resolved KR measurements in monolayer WS$_2$, along with its spectral dependence.  Like MoS$_2$, WS$_2$ also exhibits very narrow Hanle-Kerr signals and nanosecond polarization relaxation times. The peak Hanle-Kerr signal also occurs on the low-energy side of the ``A" exciton, again suggesting that coupling to resident electrons may proceed preferentially through trion transitions. The field-independent background on the Hanle data is significantly larger (about 80\% of the peak, rather than 30-40\% for MoS$_2$), which may indicate lesser material quality.

In summary, steady-state and time-resolved Kerr-rotation spectroscopy directly and unambiguously reveal the very long spin relaxation and spin coherence of resident electrons in \emph{n}-type MoS$_2$ and WS$_2$ monolayers. These nanosecond timescales exceed by 2-3 orders of magnitude typical PL recombination times. In contrast with conventional III-V and II-VI semiconductors, rapid spin depolarization of itinerant electrons in MoS$_2$ occurs in very small applied $B_y$.  Supported by a model of coupled spin-valley relaxation, these data point to a novel dephasing mechanism generated by precession about the rapid-fluctuating internal spin-orbit field that is a hallmark of TMDC materials, driven by fast inter-valley scattering.  These measurements open the door for studies of the spin- and valley- dynamics of intrinsic and resident carriers in electron- and hole-doped two-dimensional TMDC semiconductors.

\noindent
\newline
\textbf{METHODS}
\small
\newline
High-quality triangular crystals of $n$-type (electron doped) monolayer MoS$_2$ and WS$_2$ were grown by chemical vapor deposition on SiO$_2$/Si substrates following \cite{JLou}. Typical lateral dimensions ranged from 10-20 $\mu$m. The background electron doping level in these crystals is estimated to be $n_e \sim 5\times10^{12}$ cm$^{-2}$ based on transport studies of field-effect transistor devices fabricated from similarly-grown crystals. The samples were mounted on the vacuum cold finger of a small variable-temperature liquid helium optical cryostat (3-300~K).  Applied magnetic fields up to 300~mT were generated using an external electromagnet.  Individual crystals were screened for a high degree of circularly-polarized exciton PL ($>80$\%) and sharp reflectivity spectra.

All Kerr-effect measurements were performed in a reflection geometry, as depicted in Fig. 1a. For Hanle-Kerr measurements of steady-state polarization, continuous-wave (cw) pump and probe lasers were used. The 632.8~nm and 543.5~nm lines of HeNe lasers were used for the pump, while the probe laser was typically a tunable narrowband dye laser or fixed-wavelength diode lasers.  Time-resolved measurements used ultrafast 250~fs pulses from a wavelength-tunable 76~MHz optical parametric oscillator (OPO).  Time-resolved studies typically used wavelength-degenerate pump and probe pulses (an exception is the temperature-dependent data of Fig. 3, where a 635~nm pulsed diode laser with $\sim$100~ps pulse duration was synchronized to the OPO and used as a non-degenerate pump laser. In all experiments, the pump laser was weakly focused so as to illuminate the entire TMDC crystal, thereby mitigating any possible influence of density gradients or carrier diffusion, while the linearly-polarized and normally-incident probe laser was more tightly focused to a $\sim$4 $\mu$m spot in the center of the crystal. It was verified that the small off-axis angle of the pump laser (about 10 degrees from sample normal) did not influence the results -- similar data were obtained with co-propagating normally-incident pump and probe lasers.

Except where otherwise noted, the pump laser polarization was modulated between RCP and LCP by a photoelastic modulator to facilitate lock-in detection. Detection of the induced optical polarization rotation (KR) and ellipticity (KE) imparted on to the probe beam was achieved with a standard optical bridge arrangement using balanced photodiodes. Low-power pump and probe beams were used, as it was observed that the Hanle widths and polarization decay rates increased with increasing pump and/or probe power. Typical average probe power was in the tens of $\mu$W, while the pump was in the range from 100-1000 $\mu$W depending on wavelength and temperature.

\textbf{Acknowledgements}
We gratefully acknowledge D.L.~Smith and H. Dery for helpful discussions, and W.D.~Rice for laser expertise.  This work was supported by the Los Alamos LDRD program under the auspices of the US DOE, Office of Basic Energy Sciences, Division of Materials Sciences and Engineering. We also acknowledge the support from AFOSR (grant FA9550-14-1-0268) and the Welch Foundation (grant C1716).

\newpage

\section{Supplemental Information}
\subsection{Wavelength dependence of time-resolved Kerr and Hanle-Kerr measurements}
Figure S1 shows time-resolved Kerr-rotation measurements in monolayer MoS$_2$, at $B_y=$ 0 and 270~mT, using degenerate pump and probe lasers for different photon energies spanning the ``A" exciton. At short wavelengths (high photon energies), the induced KR signals are small. At intermediate wavelengths $\lambda$ just on the low-energy side of the exciton resonance (\emph{e.g.} $\lambda$=654-660~nm), the KR signals are large and decay slowly at zero field, consistent with the long-lived decay of polarized itinerant and resident electrons. At $B_y$=270 mT, the large KR signals decay quickly due to dephasing caused by precession about the combined applied field $B_y$ and the large effective internal spin-orbit field $\pm \hat z B_{\textrm{so}}$ that is rapidly fluctuating due to fast intervalley scattering of itinerant electrons.  At these wavelengths, the additional long-lived oscillatory (coherent) spin precession signal is barely visible.

\begin{figure} [b]
    \centering
      \includegraphics[width=0.5\textwidth]{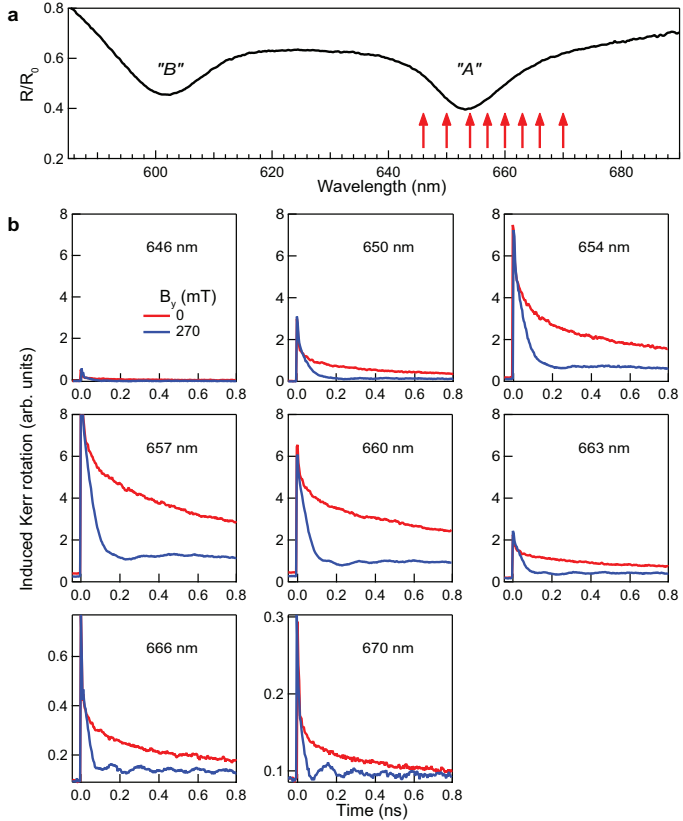}
        \caption{\textbf{Supplemental Figure S1: Wavelength dependence of time-resolved Kerr rotation studies in MoS$_2$.} \textbf{a}, Normalized reflectance spectrum $R/R_0$ from a monlayer MoS$_2$ crystal at 5~K. The arrows indicate the wavelengths $\lambda$ at which the degenerate pump-probe KR measurements were performed. \textbf{b}, Time-resolved KR in monolayer MoS$_2$ for various $\lambda$ at $B_y=$ 0 and 270~mT.  Note the changing $y$-axis scales.}\label{fig:S1}
\end{figure}

However at even longer wavelengths further on the low-energy side of the exciton resonance, the KR signals reduce until only this small oscillatory signal remains. These results are consistent with the laser pulses pumping (and probing) predominantly \emph{itinerant} electrons in MoS$_2$ at intermediate wavelengths (654-660~nm), but then pumping and probing more \emph{localized} states at longer wavelengths (663-670~nm).  Being localized, these electrons may not undergo rapid intervalley scattering and may not `see' the large spin orbit field $\pm \hat z B_{\textrm{so}}$, and therefore precess about the bare applied magnetic field $B_y$. Localized states in monolayer TMDC materials have been observed in recent experimental studies \cite{He}.

Figure S2 shows Hanle-Kerr data from monolayer MoS$_2$ at 5K.  A fixed cw pump laser was used (632.8 nm), and the different Hanle curves show data obtained with different cw probe laser wavelengths. The inversion of the induced KR signal is clearly observed (as in Fig. 2 of the main text).  These curves have somewhat greater width than those shown in the main text (\emph{e.g.}, Figure 1), due to the use of higher pump and probe laser intensity. Minimum Hanle widths (implying long spin lifetimes of resident electrons) are obtained at long wavelengths on the low-energy side of the ``A" exciton resonance.

\begin{figure}[h]
\includegraphics[width=.45\textwidth]{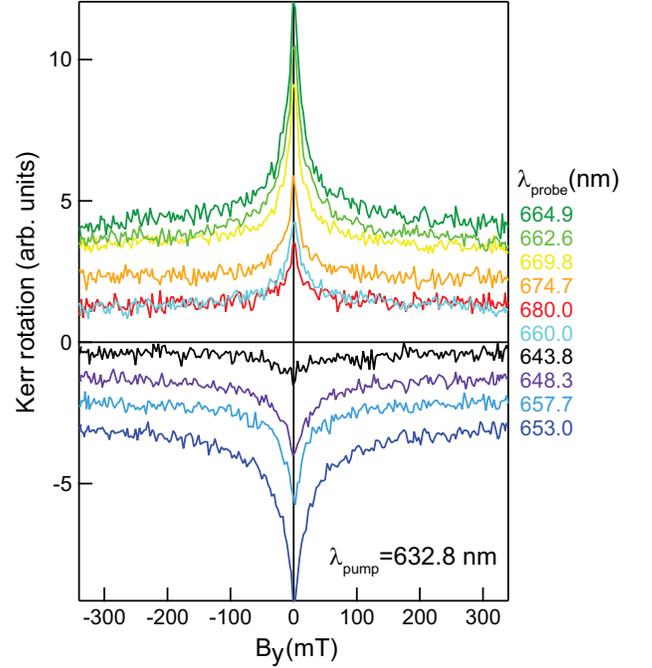}
\caption{{\bf Supplemental Figure S2: Probe wavelength dependence of Hanle-Kerr measurements in MoS$_2$}. Using cw pump and probe lasers, the plot shows the Hanle-Kerr curves measured in MoS$_2$ at 5K using different probe wavelengths spanning the ``A" exciton resonance.} \label{figS2}
\end{figure}
\newpage

\subsection{Selectively probing polarization dynamics in $K$ and $K'$ valleys}

\begin{figure}[t]
\includegraphics[width=.45\textwidth]{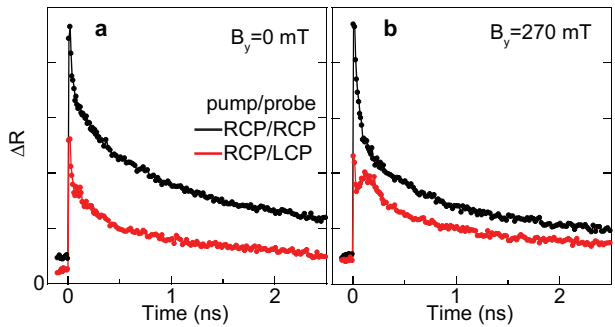}
\caption{{\bf Supplemental Figure S3: Selective right- and left-circularly polarized (RCP/LCP) pump-probe studies of induced reflectivity}. Using RCP pump pulses, the $K$ valley of monolayer MoS$_2$ is weakly photoexcited at the ``A" exciton, while the induced change in reflectivity $\Delta R$ of both RCP and LCP probe pulses is detected as a function of time.  \emph{Both} RCP and LCP probes (probing $K$ and $K'$ valleys, respectively) show an induced longlived change. $\lambda_{pump}$ = $\lambda_{probe}$ = 660~nm; $T$=5K. \textbf{a}, Measurement at $B_y$=0. \textbf{b}, Measurement at $B_y$=270~mT.} \label{figS3}
\end{figure}

Because Kerr signals scale as the \emph{difference} between RCP and LCP optical constants, it is natural to ask which component contributes primarily. That is, does resonant pumping the $K$ valley with RCP pump light affect only RCP probe light, or does it also affect LCP light (which probes the $K'$ valley)?  Figure S3 shows time-resolved measurements of pump-induced reflectivity in MoS$_2$ using co- and cross-circular pump and probe pulses. Interestingly, the reflectivity of \emph{both} RCP and LCP probe light increases immediately following photoexcitation (albeit by different amounts), consistent with recent studies on very short ($<$10 ps) timescales \cite{Mai}, suggesting that strong exciton Coulomb correlation effects while holes are present generate the initial non-equilibrium electron densities, which then, following recombination, relax on long timescales.

\end{document}